\documentclass[twocolumn,superscriptaddress,showpacs,preprintnumbers,amsmath
,amssymb,aps,prd]{revtex4}
\usepackage{graphicx}
\usepackage{bm}
\def\jstar{$J_*/M^2_{adm} \approx 0.985$}

\def\thetamax{$\theta \approx 15.2^\circ$}
\def\maxspin{$a/M_h \approx 0.99\pm 0.01$}

\def\goldenA{$\lbrace \widehat M_h/M, \hat a/M_h \rbrace = \lbrace  0.951, 0.685 \rbrace$}
\def\goldenB{$\lbrace \widehat M_h/M, \hat a/M_h \rbrace =\lbrace 0.936, 0.800 \rbrace$}
\def\goldenC{$\lbrace \widehat M_h/M, \hat a/M_h \rbrace = \lbrace  0.906\pm 0.001, 0.900\pm 0.001\rbrace$}

\def\oldaf{$a/M_h \approx 0.82$}

\def\pair{$\lbrace M_h/M, a/M_h \rbrace$}
\def\gpair{$\lbrace \widehat M_h/M, \hat a/M_h \rbrace$}

\def\ahz#1{apparent horizon#1 (AH#1)\gdef\ahz{AH}}
\def\bh#1{black hole#1 (BH#1)\gdef\bh{BH}}
\def\bbh#1{binary black holes#1 (BBH#1)\gdef\bbh{BBH}}
\def\etal{{\it et al.}}

\begin{document}

\title{Final Mass and Maximum Spin of Merged Black Holes and the Golden Black Hole}
\author{James Healy}
\affiliation{Center for Gravitational Wave Physics\\
The Pennsylvania State University, University Park, PA 16802}
\author{Pablo Laguna}
\affiliation{Center for Relativistic Astrophysics and
School of Physics\\
Georgia Institute of Technology, Atlanta, GA 30332}
\author{Richard A. Matzner}
\affiliation{Center for Relativity and Department of Physics\\
The University of Texas at Austin, Austin, TX 78712}
\author{Deirdre M. Shoemaker}
\affiliation{Center for Relativistic Astrophysics and
School of Physics\\
Georgia Institute of Technology, Atlanta, GA 30332}

\begin{abstract} 
  We present results on the mass and spin of the final black hole from
  mergers of equal mass, spinning black holes. The study extends over
  a broad range of initial orbital configurations, from direct plunges
  to quasi-circular inspirals to more energetic orbits
  (generalizations of Newtonian elliptical orbits). It provides a
  comprehensive search of those configurations that maximize the final
  spin of the remnant black hole. We estimate that the final spin can
  reach a maximum spin \maxspin ~for extremal black hole mergers. In
  addition, we find that, as one increases the orbital angular
  momentum from small values, the mergers produce black holes with
  mass and spin parameters \pair ~spiraling around the values \gpair
  ~of a {\it golden} black hole. Specifically, $(M_h-\widehat M_h)/M
  \propto e^{\pm B\,\phi}\cos{\phi}$ and $(a-\hat a)/M_h \propto
  e^{\pm C\,\phi}\sin{\phi}$, with $\phi$ a monotonically growing
  function of the initial orbital angular momentum. We find that the
  values of the parameters for the \emph{golden} black hole are those
  of the final black hole obtained from the merger of a binary with
  the corresponding spinning black holes in a quasi-circular inspiral.
\end{abstract}

\maketitle

%%%%%%%%%%%%%%%%%%%%%%%%%%%%%%%%%%%%%%%%%%%%%%%%%%%%%%%%%%%%%%%%%%%%%%%%%%%%

\begin{figure}
 \includegraphics[width=6.0cm,angle=270]{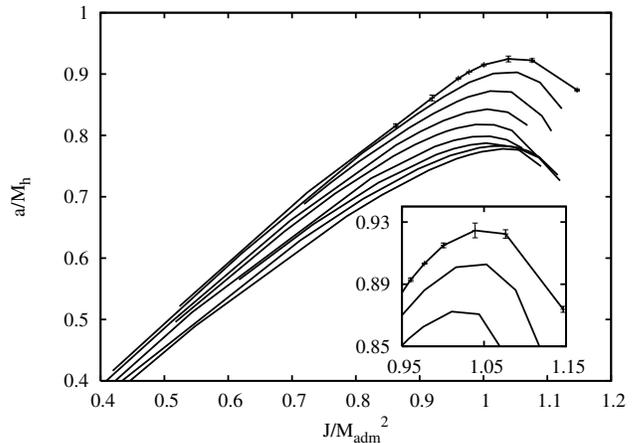}
 \caption{Final spin $a/M_h$ {\it vs} the initial total (orbital plus
   spin) angular momentum $J/{M_{adm}}^2$ at a constant
   $\theta=26.565^{\circ}$. From top to bottom are the cases $S/M^2 =
   \{0.4,0.3,0.2,0.1,0.0,-0.1,-0.2,-0.3,-0.4\}$, respectively. As a
   reference, error bar estimates are given for the $S/M^2 = 0.4$
   case. }
\label{fig:spinJ}
\end{figure}

\emph{Introduction:} An important question that can only be answered
with the tools provided by numerical relativity is: What is the final
mass and maximum spin of the final \bh{} from generic binary mergers?
In Ref.~\cite{2008PhRvL.101f1102W}, referred to here as Paper I, we
took the first step towards answering this question. We studied the
behavior of a $1-$parameter series of equal mass, non-spinning \bh{s}
in merger or fly-by. The adjustable parameter was the magnitude of
initial linear momentum $P$ for each of the \bh{s}. We kept fixed the
initial binary coordinate separation $\vec d=10\,M\,\hat x$ and the
angle $\theta=26.565^{\circ}$ between $\vec P$ and $\vec d$. That is,
the binaries had initial orbital momentum $\vec{L} =
d\,P\,\sin{\theta}\, \hat z= 4.47\,M\,P\, \hat z$ and impact parameter
$b = d\,\sin{\theta} = 4.47\,M$. Here $M=2\,m$ is the total mass of
the binary, where $m$ is the mass of each \bh{.} The main result in
Paper I was the specific dependence or transfer function we found that
connects the final merged \bh{} spin $a/M_h$ with the initial
$L/M^2_{adm}$ (see Fig.~1 in Paper I, which is repeated in
Fig.~\ref{fig:spinJ} here in the curve labeled as $S/M^2= 0.0$,
\emph{i.e.} vanishing total initial spin). For this case, a maximum
final \bh{} spin \oldaf ~was found at $L/M^2_{adm} \approx 1$.

The result of Paper I has particular astrophysical interest as an
upper limit on the spin from initially non-spinning \bh{} mergers,
though \bh{s} can be spun up by other processes as well, specifically
by accretion. No previous \bh{} merger simulation has produced final
spins close to maximal. We do know from specific examples in work by
Campanelli \emph{et al.} \cite{2006PhRvD..74d1501C} that ``hangups''
occur, in which the orbital radius decreases only slowly or not at
all, and which enable the radiation of substantial angular momentum
before the merger. In Paper I, we also found a splash-skip behavior
involving what Pretorius and Khurana \cite{2007CQGra..24...83P} called
whirl orbits. In these cases the interacting \bh{s} can approach, then
recede to a large distance, then approach again, producing repeated
bursts of gravitational radiation. This is clearly an extreme case of
``hangup.''

Here we extend the work in Paper I to encounters where at least one of
the \bh{s} carries a spin, with the spins parallel or anti-parallel to
the orbital angular momentum. The main motivation is to add extra
initial angular momentum in order to increase the final \bh{} spin.
In addition, we vary the angle $\theta$, but keep the initial
separation as before, $d=10\,M$. We find similar transfer functions to
that in Paper I, with the maximum final \bh{} spin depending on the
spins of the merging \bh{s} and the scattering angle $\theta$. Our
simulations show that the final \bh{} can reach a maximum spin
\maxspin ~when extrapolated to the merger of maximally spinning \bh{s}.

In addition, we found an interesting behavior regarding the parameters
\pair ~of the final \bh{} as we increased the orbital angular
momentum. The values for \pair ~spiral around the parameters of what
we operationally call a {\it golden} \bh{.} We have been able to
identify the \emph{golden} \bh{} to be the hole obtained from the
merger of a binary in a quasi-circular inspiral with the corresponding
spinning \bh{s}.

The computational setup for the simulations presented here is very
similar to the one in Paper I, the main difference being the spins of
the colliding \bh{s}. All simulations had 10 levels of refinement,
sixth order spatial differencing, with an outer boundary of $\sim
320\,M$. For the runs used to compute the $\theta \geq 26.565$ maximum
final spin ($\sim 200-300\,M$ runtimes) a resolution of $M/51.6$ was
used on the finest level, with each subsequent level decreased by a
factor of 2. For the $\theta < 26.565^{\circ}$ runs ($\sim 200-300\,M$
runtimes), a resolution of $M/103$ was used.  For the longer runs to
compute the \emph{golden} \bh{} ($\sim 500-3000\,M$ runtimes), a
resolution of $M/90$ was used on the finest grid.

\begin{figure} \begin{center}
   \includegraphics[width=6.0cm,angle=270]{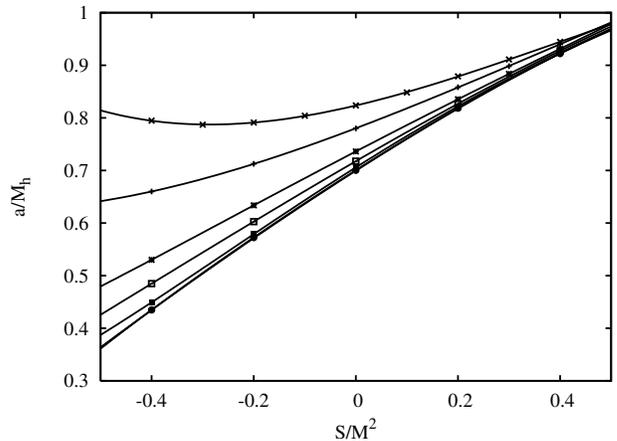}
 \end{center} \caption{Maximum final spin
   {\it vs} the initial spin angular momentum $S/M^2$. From top to
   bottom are the cases $\theta =
   \{26.565^\circ,32^\circ,45^\circ,55^\circ,70^\circ,80^\circ,90^\circ\}$,
   respectively.}
\label{fig:MaxSpin}
\end{figure}

\begin{figure} \begin{center}
   \includegraphics[width=5.5cm,angle=270]{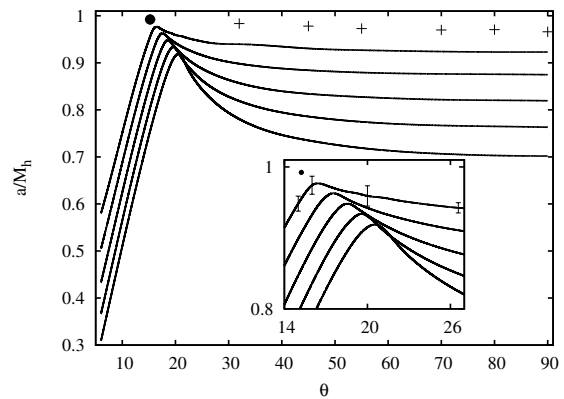}
 \end{center} \caption{Maximum final spin
   {\it vs} $\theta$ for $S/M^2 = 0.0, 0.1, 0.2, 0.3, 0.4$ (bottom to top)}
\label{fig:MaxSpinTheta}
\end{figure}

The present study required $\sim 1,000$ \bbh{} simulations. Every data
point displayed in each figure involves at least one \bbh{} simulation
and in Figs.~\ref{fig:MaxSpin} and \ref{fig:MaxSpinTheta} up to ten
simulations.  It was thus prohibitively expensive to do the runs
needed to estimate errors via Richardson extrapolation for each data
point. With this in mind, we chose to do all the simulations at the
resolutions mentioned in the previous paragraph, which by experience
we know our code yields sufficiently accurate results.  In order to
get error estimates, we selected a subset of representative cases, in
some instances the most challenging cases (e.g. high \bh{} spins), to
perform convergence tests and apply Richardson extrapolation.  Those
are the cases with error bars in Figs.~\ref{fig:spinJ},
\ref{fig:MaxSpinTheta} and \ref{fig:gold_error}. The convergence tests
demonstrated consistency with the expected fourth order accuracy of
our code. In addition, the Richardson extrapolation of results yielded
error bars for the spin and mass of the final black hole $\sim
0.1\%$. For situations involving \bh{} spins $a/M_h \ge 0.95$, we
identified additional sources of error due to inaccuracies in finding
the exact location of the \ahz{} that yield errors in the range
between 0.5 and $1\%$.

%%%%%%%%%%%%%%%%%%%%%%%%%%%%%%%%%%%%%%%%%%%%%%%%%%%%%%%%%%%%%%%%%%%%%%%%%%%%

\emph{Maximum Final Black Hole Spin:} The first component of this work
is aimed at addressing the following three questions regarding the
maximum spin of the final \bh{:} Does the maximum occur near the same
initial angular momentum as found in Paper I? What is the dependence
of the maximum residual spin on the value of the initial spins of the
merging \bh{s}? How close is the maximum final spin to the Kerr limit?
The answer to the first question is ``yes''. The maximum for the
residual angular momentum occurs near $J/M^2_{adm} \approx 1$ for all
cases considered. This conclusion is apparent from
Fig.~\ref{fig:spinJ}, where we show the dependence of the spin of the
final \bh{} $a/M_h$ as a function of the initial total (orbital plus
spin) angular momentum $J/M^2_{adm}$ at a constant
$\theta=26.565^{\circ}$. From top to bottom are the cases $S/M^2 =
\{0.4,0.3,0.2,0.1,0.0,-0.1,-0.2,-0.3,-0.4\}$, respectively, where
$S/M^2 = (ma_1 +ma_2)/(2m)^2=(a_1/m+a_2/m)/4$ is the total initial
spin of the binary with $a_1/m$ and $a_2/m$ the individual \bh{}
dimensionless spin parameters. For reference, in the $S/M^2 = 0.4$
case, we show error bar estimates obtained from Richardson
extrapolation.

An interesting feature in the results displayed in
Fig.~\ref{fig:spinJ} is the crossing of lines for negative
$S/M^2$. The crossing occurs because the mergers with negative $S/M^2$
do not suffer the ``hangup'' of the positive
counterparts~\cite{2006PhRvD..74d1501C}. Therefore, as $S/M^2$ becomes
more negative, one is able to increase the total angular momentum
$J/M^2_{adm}$ to higher values before reaching the maximum spin
$a/M_h$, which roughly corresponds to the transition between direct
plunges and inspiral-like coalescences.

We have looked for but find very little torque-up, suggesting very
little spin-orbit coupling. At the same time, for the configurations
we consider, the final spin as a function of $J$ depends only on the
total initial spin, not how it is distributed between the interacting
\bh{s}. For instance, the cases $a_{1,2}/m = \{0.4,0.4\}$
($S/M^2=0.1+0.1$), $a_{1,2}/m= \{0.0,0.8\}$ ($S/M^2=0.0+0.2$) and
$a_{1,2}/m= \{0.6,0.2\}$ ($S/M^2=0.15+0.05$), which have the same
initial total spin $S/M^2 = 0.2$ but distributed differently between
the two \bh{s}, produce almost identical final merged \bh{} angular
momentum $vs.$ the initial total angular momentum $J/M_{adm}^2$.  We
have carried out similar experiments for total $S/M^2 = 0.1$ (with
$a_{1,2}/m= \{0.8,-0.4\}$ and $a_{1,2}/m= \{0.2,0.2\}$) and $S/M^2 =
0$ (with $a_{1,2}/m= \{0.0,0.0\}$, $a_{1,2}/m= \{0.4,-0.4 \}$ and
$a_{1,2}/m= \{0.8,-0.8\}$). These results suggest that the initial
spin is simply ``bundled into" the total $J$ in the evolution. A
fraction of $J$ is then deposited into the final \bh{,} so that the
final \bh{} spin depends essentially only on the {\it total} initial
angular momentum.

This result is confounding because a close examination of the orbits
shows significant differences between two cases with the same total
but different distribution of initial spin. For instance, the case
$a_{1,2}/m= \{0.4,0.4\}$ with symmetrical initial data yields a
significantly different evolution from the case $a_{1,2}/m =
\{0.0,0.8\}$. In fact, the $a_{1,2}/m=\{0.0,0.8\}$ case produces a
$kick$ of $\sim 175$ km/sec, and the symmetric case $a_{1,2}/m=
\{0.4,0.4\}$ produces no kick~\cite{2007PhRvD..76h4032H,2007ApJ...661..430H,
  Koppitz:2007ev,2008ApJ...679.1422R}.

We argued in Paper I that the existence of the maximum in the final
spin parameter $a/M_h$ depends on the participation of an
``intermediate excited state'' which we characterized as essentially a
highly distorted \bh{.} Existence of such an intermediate state seems
to be generic, and it emits the largest part of the radiated energy
and angular momentum. We conjecture that the non-linearity of such a
state will produce such strong radiation of angular momentum that it
will enforce an upper limit on the residual \bh{} angular momentum. We
extend that argument here to claim that, while in this state,
substantial information is lost about the initial configuration of the
system.

The answers to the second question (``how does the final \bh{} depend
on the initial total spin?") can be found in Fig.~\ref{fig:MaxSpin}
where we show the maximum spin $a/M_h$ deposited in the final \bh{}
{\it vs} the initial total spin angular momentum $S/M^2$. From top to
bottom are $\theta =
\{26.565^\circ,32^\circ,45^\circ,55^\circ,70^\circ,80^\circ,90^\circ\}$,
respectively.  Each $\theta = $ constant line in Fig.~\ref{fig:MaxSpin}
is obtained from the set of maximum final spins like those found in
Fig.~\ref{fig:spinJ} for $\theta=26.565^{\circ}$.  Notice in
Fig.~\ref{fig:MaxSpin} that the dependence of the final spin on the
spins of the merging \bh{s} is almost linear for $\theta$ approaching
$90^\circ$ but non-trivial for small values of $\theta$. The small
$\theta$ values correspond to small impact parameters. For these
cases, mergers with negative $S/M^2$ are direct plunges; very little
angular momentum is radiated, thus the turn around observed in
Fig.~\ref{fig:MaxSpin}.

A possible answer to the third question (``what is the $maximum$ final
spin in these mergers?") can be found by extrapolating the $\theta = $
constant lines in Fig.~\ref{fig:MaxSpin} to maximally spinning
incident \bh{s} ($i.e.$ to $S/M^2 = 0.5$). Such extrapolation yields a
final \bh{} spin $a/M_h \approx 0.98$. This maximum final spin is
slightly lower than the prediction for quasi-circular inspirals by
Kesden~\cite{2008PhRvD..78h4030K} of $a/M_h \approx 0.9988$ but higher
than the estimate by Rezzolla~\cite{2009CQGra..26i4023R} of $a/M_h
\approx 0.959$. Recently, work by Sperhake
\etal~\cite{Sperhake:2009jz} has also produced final spins $a/M_h
\approx 0.95$ for non-spinning, highly boosted \bh{s}. We believe that
these highly boosted \bh{s} address a different regime from the one
considered here. The mergers in Sperhake \etal~\cite{Sperhake:2009jz}
involve ultra-relativistic encounters with initial total angular
momentum as high as $J/M^2_{adm} \approx 3$ and small enough impact parameter to
``force'' a merger; that is, the binaries are initially
``over-saturated'' with angular momentum. In our case, we start with
$J/M^2_{adm} \approx 1$ configurations and arrange the binary to
minimize the radiation losses of angular momentum.

Another important aspect to keep in mind regarding the search for the
maximum final spin is that our study involves a 2-parameter family of
simulations in $S/M^2$ and $\theta$.  Extrapolation along $\theta = $
constant as in Fig.~\ref{fig:MaxSpin} may not be the right path in
parameter space to find the maximum final spin. To further investigate
this point, we plot in Fig.~\ref{fig:MaxSpinTheta} the final spin as a
function of $\theta$ but now for lines of $S/M^2 = $ constant. From
bottom to top are the cases $S/M^2 = 0.0, 0.1, 0.2, 0.3, 0.4$. The
crosses along the top axis of the figure are the values obtained from
the $\theta = $ constant extrapolation to extremal \bh{s} in
Fig.~\ref{fig:MaxSpin}.  It is now clear why the maximum final spin
values obtained from extrapolating the data in Fig.~\ref{fig:MaxSpin}
are similar to each other (i.e. all lines approach each other for
$S/M^2 = 0.5$). As Fig.~\ref{fig:MaxSpinTheta} shows, the final spin
for $\theta \ge 30^\circ$ along $S/M^2 = $ constant lines is roughly
constant thus yielding similar extrapolation values along $\theta = $
constant (see crosses in Fig.~\ref{fig:MaxSpinTheta}).

Notice that in Fig.~\ref{fig:MaxSpinTheta} we have extended the range
of simulations to include $\theta \le 26.565^\circ$. A completely
different $\theta$ dependence is found for these smaller angles, whose
details are better appreciated in the inset of
Fig.~\ref{fig:MaxSpinTheta}, where we have also included the error
bars estimates for the $S/M^2 = 0.4$ case. As $\theta$ decreases along
each $S/M^2 = $ constant line, the final spin increases and reaches a
maximum. Notice in particular from the inset
Fig.~\ref{fig:MaxSpinTheta}, that the top line, $S/M^2 = 0.4$ case, is
not as smooth as the others due to an artificial loss of angular
momentum like that seen by Marronetti
\etal~\cite{Marronetti:2007wz}. We however are able to obtain a
maximum spin of $a/M_h = 0.98 \pm 0.01$ around $\theta = 16^\circ$.
Notice then from Fig.~\ref{fig:MaxSpinTheta} that a better estimate of
the maximum (and hopefully global) final spin can be found by an
extrapolation to extremal \bh{s} using the values of the maximum on
each the $S/M^2 = $ constant lines in Fig.~\ref{fig:MaxSpinTheta}. The
result of this extrapolation yields a \emph{maximum final spin}
\maxspin ~at \thetamax ~and is depicted as a black dot in
Fig.~\ref{fig:MaxSpinTheta}. Also very important to point out is that,
although this study involved simulations in the highly non-linear
regime of general relativity, the data used to extrapolate to maximum
final spin exhibit only a slight departure from linearity (see maximum
values in the inset of Fig.~\ref{fig:MaxSpinTheta} and upper right
corner data in Fig.~\ref{fig:MaxSpin}).  In summary, our study
consists of a 2-parameter $\lbrace S/M^2, \theta \rbrace$ family of
simulations, with global maximum \maxspin{} of the final spin found by
slicing the data along $S/M^2 = $ constant lines.

%%%%%%%%%%%%%%%%%%%%%%%%%%%%%%%%%%%%%%%%%%%%%%%%%%%%%%%%%%%%%%%%%%%%%%%%%%%%

\begin{figure} \begin{center}
   \includegraphics[width=5.5cm,angle=270]{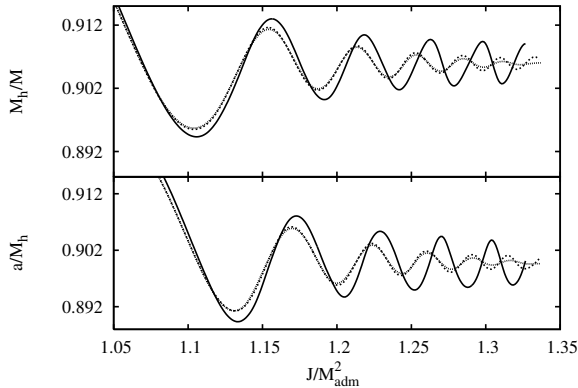}
 \end{center} \caption{Final mass $M_h/M$ (top panel) and spin 
   $a/M_h$ (bottom panel) {\it vs} $J/M_{adm}^2$ for
   $\theta = 60^\circ \,(\hbox{solid}), 80^\circ \,(\hbox{dash})$ and
   $90^\circ\,(\hbox{dotted})$ with initial total spin of $S/M^2 = 0.4$.
   The $80^\circ$ and $90^\circ$ results almost exactly overlap.}
\label{fig:AMosc} \end{figure}

\begin{figure} \begin{center}
   \includegraphics[width=5.5cm,angle=270]{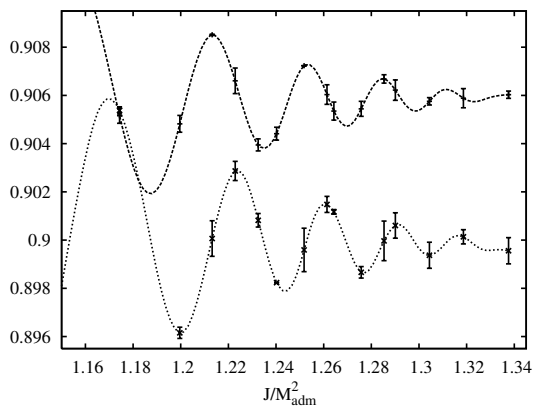}
 \end{center} \caption{Final mass $M_h/M$ (dashed, top) and spin
   $a/M_h$ (dotted, bottom) {\it vs} $J/M_{adm}^2$ with error bars for
   $\theta = 90^\circ$ and $S/M^2 = 0.4$.}
\label{fig:gold_error} \end{figure}

\emph{Golden Black Hole:} In addition to studying the maximum possible
final spin, we investigated the final state \pair ~as one increases
the initial angular momentum $J/M_{adm}^2$. The results for the case
$S/M^2 = 0.4$ with $(a_{1,2}/m=0.8)$ and angles $\theta = \{60^\circ,
80^\circ,90^\circ\}$ can be found in Fig.~\ref{fig:AMosc}.  In order
to get a sense of the accuracy of our results, we repeat in
Fig.~\ref{fig:gold_error} the case $\theta = 90^\circ$ and $S/M^2 =
0.4$ including error bar estimates for both the final mass and spin.
It is interesting that as $J/M_{adm}^2$ is increased, the pair \pair
~exhibits decaying oscillations around some values \gpair. After a
close examination of the orbits that lead to these results, we
concluded that these oscillations are a direct consequence of the
binary having to undergo multiple zoom-whirl episodes to shed the
``excess'' of angular momentum that inhibits the
merger~\cite{Healy:2009zm}. Note also that for a given angle $\theta$,
the amplitude of the oscillations are closely similar between $a/M_h$
and $M_h/M$. The two quantities are $90^\circ$ out of phase.
Furthermore, the frequency of the oscillations is not
constant but increases for larger $J/M_{adm}^2$. 

\begin{figure} \begin{center}
 \includegraphics[width=5.5cm,angle=270]{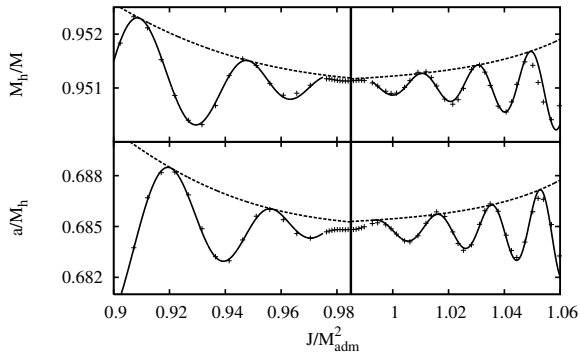}
\end{center} \caption{$M_h/M$
(top) and $a/M_h$ (bottom) {\it vs} $J/M_{adm}^2$ for $\theta = 90^\circ$
with initial total spin of $S/M^2 = 0.0$.} \label{fig:fit}
\end{figure}

In order to get a better insight on the oscillation and, in
particular, to investigate whether the damping of the oscillations
continues for larger $J/M_{adm}^2$, we focused our attention on the
case $S/M^2 = 0$. Vanishing initial spins not only yield shorter
merger times but also require lower resolutions, thus reducing the
computational cost of the large number of simulations needed.
Figure~\ref{fig:fit} shows $M_h/M$ (top) and $a
/M_h$ (bottom) {\it vs} $J/M_{adm}^2$ for $\theta = 90^\circ$ and
non-spinning initial \bh{s}. Crosses are the data from the simulations
and solid lines are fits to
\begin{eqnarray}
\frac{a}{M_h} &=& \frac{\hat a}{M_h} + A_\pm\,e^{\pm B_\pm\,\phi}\sin(\phi)\\
\frac{M_h}{M} &=& \frac{\widehat M_h}{M} + D_\pm\,e^{\pm C_\pm\,\phi}\cos(\phi)
\end{eqnarray}
with $\phi$ a monotonically growing third order polynomial in
$J/M_{adm}^2$. The fittings were done as follows: Around \jstar
~(vertical line in Fig.~\ref{fig:fit}) the oscillation is almost
completely damped out. We have then divided the data into values below
and above $ J_*/M_{adm}^2$. For those below, we fit a decaying
exponential and, similarly, a growing exponential for the data above
values $J_*/M_{adm}^2$. For the fitting, we have ignored the four
points in the neighborhood of $J_*/M_{adm}^2$ since in that region the
oscillations are almost completely damped.

An interesting finding is that, modulo a phase shift, $\phi$ is
basically the same fitting function (i.e. similar constant
coefficients in the third order polynomial in $J/M_{adm}^2$) for the
decaying and growing sectors. We also obtain that both the decaying
and growing fittings yield the same values of \goldenA. The only
changes are in the constants $\lbrace A_\pm, B_\pm, C_\pm, D_\pm
\rbrace$. Finally, the exponential envelopes (dashed lines in
Fig.~\ref{fig:fit}) cross approximately at $ J_*/M_{adm}^2$. For this
non-spinning situation, we repeated similar analysis for $\theta =
70^\circ$ and found the same values for \gpair ~within our numerical
errors. In addition, we considered total initial spins cases $S/M^2 =
0.2$ ($\theta = 90^\circ$) and $S/M^2 = 0.4$ ($\theta = 60^\circ,
80^\circ, 90^\circ$). The $S/M^2 = 0.2$ case yielded \goldenB, and the
$S/M^2 = 0.4$ values\goldenC, with the last case selected for error
estimates. We name the \bh{s} with parameters \gpair ~the
\emph{golden} \bh{s}.

\begin{figure} \begin{center}
 \includegraphics[width=5.5cm,angle=270]{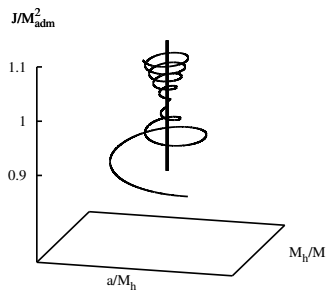}
\end{center} \caption{Final spin $a/M_h$  {\it vs} final mass $M_h/M$ {\it vs} $J/M_{adm}^2$ for
   $90^\circ$ with initial total spin of $S/M^2 = 0.0$.} \label{fig:spiral}
\end{figure}

Figure~\ref{fig:spiral} combines the results from Fig.~\ref{fig:fit}.
The vertical line denotes the location of the \emph{golden} \bh{.}
The result is a logarithmic spiral which approaches the \emph{golden}
\bh{} as $J/M_{adm}^2$ nears \jstar ~from above or below. A similar
spiral is found in the $a/M_h$ {\it vs} $M_h/M$ plane when the angular
momentum $J/M_{adm}^2$ is held constant and one plots them instead as
a function of the angle $\theta$. The center of the spiral is again at
the \emph{golden} \bh{.}  A closer look at our simulations reveals the
identity of the \emph{golden} \bh{} as the final \bh{} obtained from
the merger of a binary of the corresponding spinning \bh{s} in a
quasi-circular inspiral. We are currently investigating the reasons
behind this finding.

%%%%%%%%%%%%%%%%%%%%%%%%%%%%%%%%%%%%%%%%%%%%%%%%%%%%%%%%%%%%%%%%%%%%%%%%%%%%

\emph{Conclusions:} From the first studies of binary \bh{}
mergers~\cite{Pretorius:2005gq,2006PhRvL..96k1102B,2006PhRvL..96k1101C},
it became apparent that merger waveforms have a simpler structure than
was expected. Here we have presented two more examples of reduction of
complexity. One is the invariance of the final \bh{} spin {\it vs} the
initial total angular momentum with respect to how the individual
spins are distributed between the interacting \bh{s}. The other
invariance is the final \gpair ~of the \emph{golden} \bh{} with
respect to the angle $\theta$, for a given total initial spin $S/M^2$
(see Fig.~\ref{fig:AMosc}). These are explicit reductions of degrees
of freedom from the initial data to the final \bh{.}  From the results
of our 2-parameter family of simulations, we estimated that the
maximum final spin is found for initial extremal \bh{s} with spins
aligned with the orbital angular momentum and \thetamax. Our results
imply a maximum final spin \maxspin.

%%%%%%%%%%%%%%%%%%%%%%%%%%%%%%%%%%%%%%%%%%%%%%%%%%%%%
This work was supported in part by NSF grants PHY-0653443 (DS),
PHY-0914553 (PL), and PHY-0855892 (PL,DS). Computations carried out
under LRAC allocation MCA08X009 (PL,DS) and at the Texas Advanced
Computation Center, University of Texas at Austin. We thank Y.~Chen,
M.~Kesden and U.~Sperhake for their comments and suggestions.

\end{document}